\documentclass{article}

\usepackage{arxiv}

\usepackage[utf8]{inputenc} % allow utf-8 input
\usepackage[T1]{fontenc}    % use 8-bit T1 fonts
\usepackage{hyperref}       % hyperlinks
\usepackage{url}            % simple URL typesetting
\usepackage{booktabs}       % professional-quality tables
\usepackage{amsfonts}       % blackboard math symbols
\usepackage{nicefrac}       % compact symbols for 1/2, etc.
\usepackage{microtype}      % microtypography
\usepackage{lipsum}		% Can be removed after putting your text content
\usepackage{graphicx}
\usepackage{doi}
\usepackage[tight]{units}
\def \be {\begin{equation}}
\def \ee {\end{equation}}

\title{Topologically protected $\pi$-ring qubits}

\date{} 					% Or removing it

\author{ \href{https://orcid.org/0000-0003-3442-4445}{\includegraphics[scale=0.06]{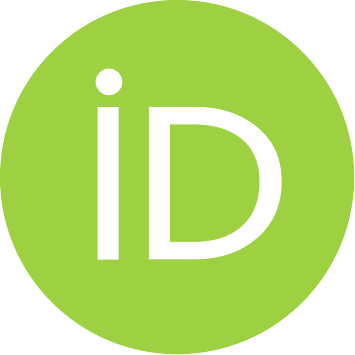}\hspace{1mm}Michael Forrester}\\
	Advanced Materials and Devices\\
	QinetiQ\\
	Farnborough\\ 
	UK\\
	GU14 0LX
	\And
	Fedor Kusmartsev \\
	College of Art and Science\\
	Khalifa University\\
	Abu Dhabi\\
	UAE\\
	P.O. Box 127788\\
	}
\begin{document}
\maketitle

\begin{abstract}
The $\pi$-ring qubit array is described using quasiclassical approaches that are shown to be accurate and give clarity to the complex energy landscape of connected vortex qubits. Using the techniques, large arrays of Josephson junction systems can be designed, including phase shift devices. Herein, connected arrays of loops containing $\pi$ junctions are described. These techniques are useful for design of quantum computers based on superconducting technologies, hybrid quantum technologies and quantum networks. 
\end{abstract}

% keywords can be removed

In topological quantum computation it is a design stipulation that the utmost protection of the qubits from the effects of environmental noise should ensue. Extraneous noise or circuitry can destroy the quantum state and drastically affect coherence times. One way of protecting the qubits is to enclose them in a medium of almost identical structures. Indeed, this gives a quantum error correction capability to the systems such that the effects of local noise become exponentially small as the size of the lattice increases \cite{IoffeFeigelman2002,IoffeFeigelman2003}. The enclosure of a central plaquette is the focus of this next investigation. We also address the inclusion of a   $\pi$- shift into each element of the array.  The superconducting   $\pi$- ring has enormous potential for sensitive field detection, energy transfer as well as quantum computation.

\begin{figure}[!htp]
\begin{center}
\includegraphics[width=16cm,keepaspectratio]{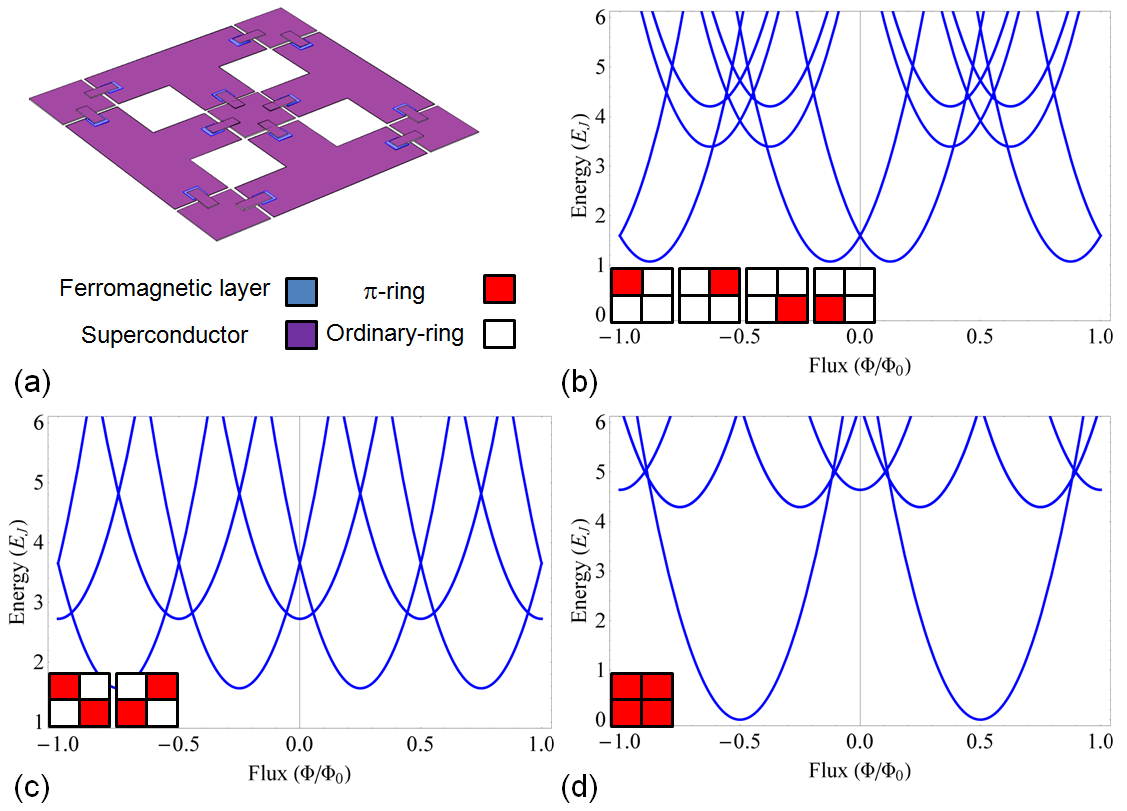}
\end{center}
\caption{A system composed of superconductor-ferromagnet-superconductor Josephson-junctions \cite{Frolov2008} with four plaquettes. The total energy is normalised by the Josephson energy as a function of the magnetic flux in the system. The energy is calculated using the quasiclassical method \cite{Kusmartsev1998}. }
\label{Fig:PiSquares}   
\end{figure}

Every interconnection in a superconducting island network contains a Josephson junction. We introduce a phase shift of $\pi$ to some of the junctions. Technologically, this can be created by linking superconducting islands through a ferromagnetic layer of the correct thickness. Frolov and co-workers did just that when determining the spontaneous currents in superconducting arrays of  $\pi$-junctions\cite{Frolov2008}. For example, they considered a two dimensional lattice of Josephson junctions, as illustrated in Fig. \ref{Fig:PiSquares} $(a)$ on the left-hand side.

\begin{figure}[!htp]
\begin{center}
\includegraphics[width=8.5cm,keepaspectratio]{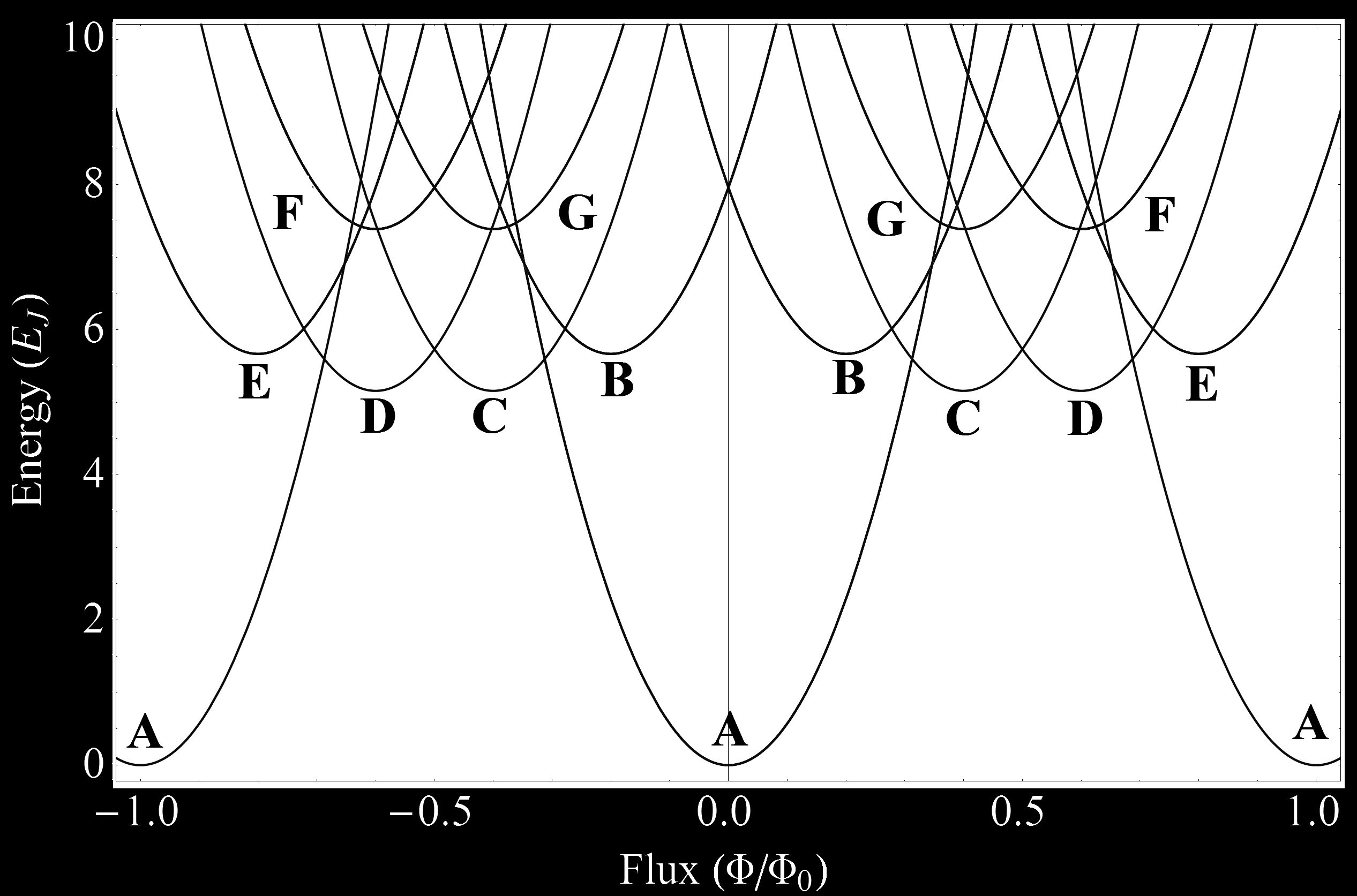}
\end{center}
\caption{The energy distribution as a function of the applied flux for a system of triangular geometries containing only ordinary junctions.}
\label{Fig:evsftriangular}   
\end{figure}

\begin{figure}[!htp]
\begin{center}
\includegraphics[width=8.5cm,keepaspectratio]{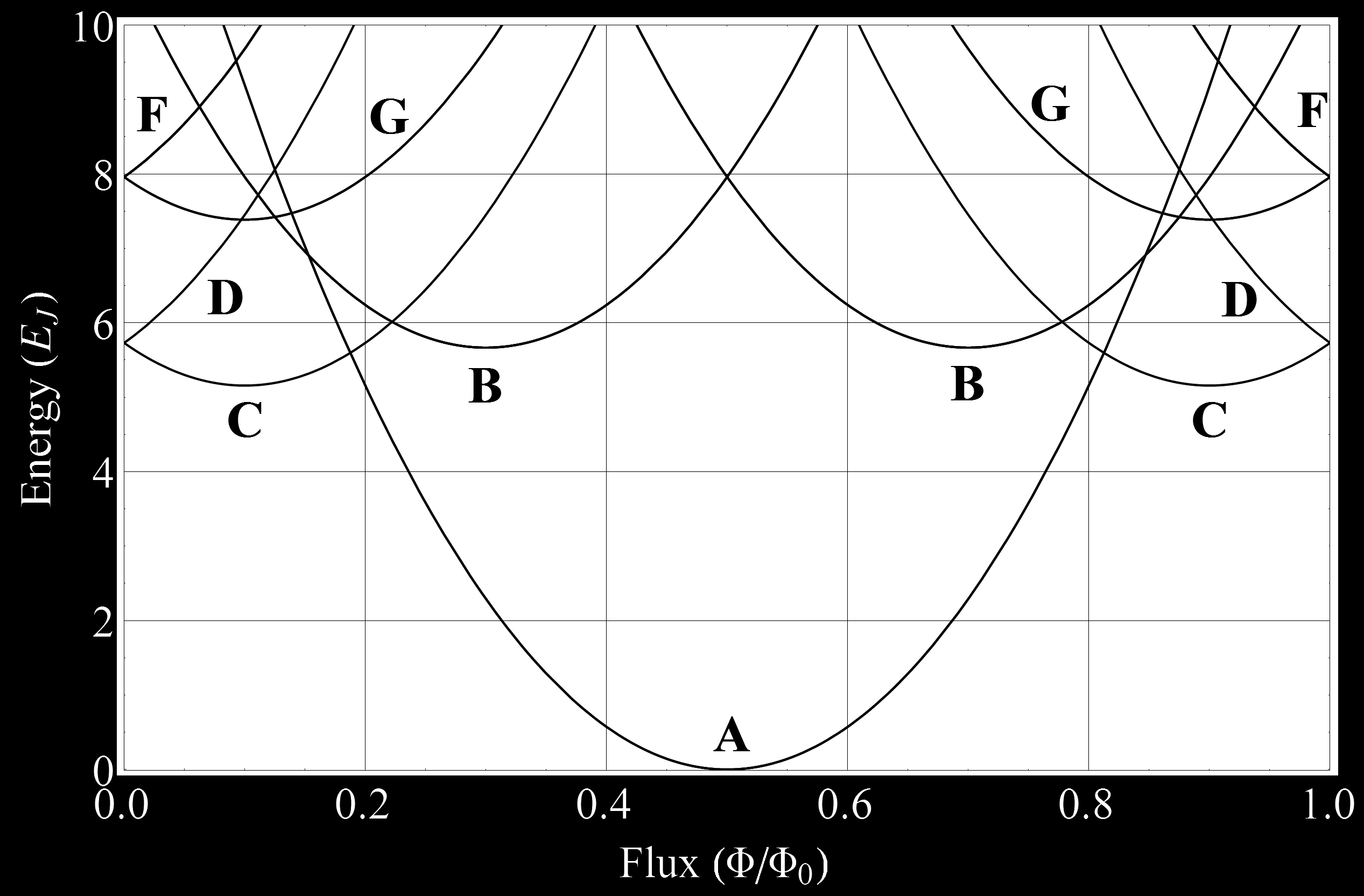}
\end{center}
\caption{The energies for the triangular Josephson junction triangular array as a function of flux. A $\pi$-shift is developed due to the introduction of an odd number of $\pi$-junctions in each plaquette.}
\label{Fig:evsftriangularpi}   
\end{figure}

\begin{figure*}[!htp]
\begin{center}
\includegraphics[width=12cm,keepaspectratio]{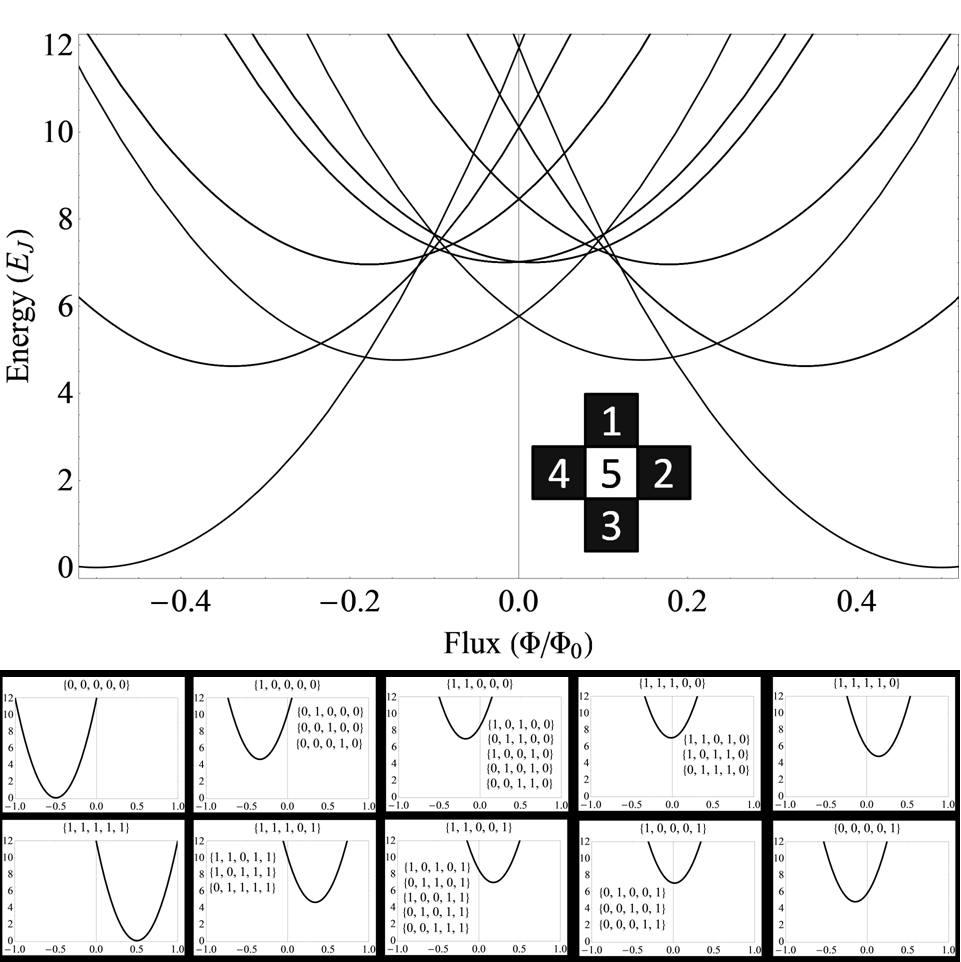}
\end{center}
\caption{The energies associated with five square plaquettes arranged in a spin star when each contains an odd number of $\pi$-junctions.}
\label{Fig:spinstar}   
\end{figure*}

In the experiments of Frolov et al\cite{Frolov2008} their objective was to show the emergence of spontaneous currents in  square superconductor-ferromagnet-superconductor ($SFS$) $\pi$-junction circuits. To do so they created arrays with different levels of frustration. They showed unequivocally that the  $\pi$-junctions generated the sought after currents and explained the spontaneous current distributions by simulating the Josephson phase dynamics. We use a quasiclassical approach and find good comparison to their findings in the energy-flux dependencies. For example, in Fig. \ref{Fig:PiSquares} $(c)$ the energy as a function of the magnetic flux per cell is shown for the case where the plaquettes on the main diagonal are  $\pi$-rings (containing odd numbers of  $\pi$-junctions) and those on the off-diagonal contain two or four  $\pi$-junctions. This has the effect of creating an antiferromagnetically aligned system. The results mirror the array free energies found in reference\cite{Frolov2008}. Thus, we have confidence that the quasiclassical method is accurate and can find all the possible ground and excited states and to resolve all the vortex states. This is in contrast to, say, Monte Carlo methods that tend to miss many of the pertinent energy levels, and frequently can only determine the two lowest energy states closest to the ground state energy\cite{Kusmartsev1998}. 
\begin{figure}[!htp]
\begin{center}
\includegraphics[width=8.5cm,keepaspectratio]{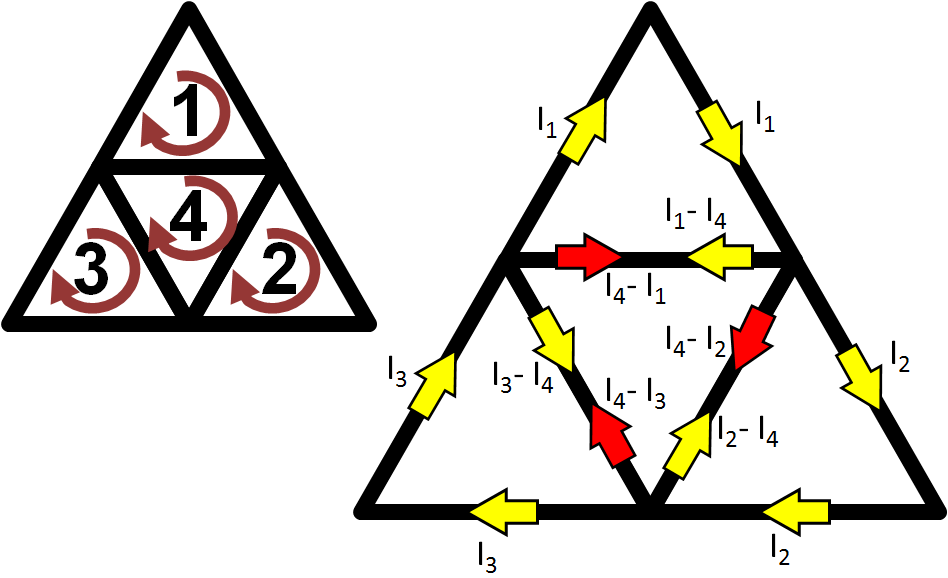}
\end{center}
\caption{Triangular plaquettes with each leg containing a Josephson junction. In the case of $\pi$-rings each plaquette has one side containing a $\pi$-junction. The currents are taken to flow in a clockwise direction around each cell and replace the phases over each junction in the harmonic approximation.}
\label{Fig:triangles}   
\end{figure}
We now demonstrate a system of four interconnected triangular plaquettes, as shown in Fig.\ref{Fig:triangles}. Such systems, that exploit geometrically induced frustration, have been made from Josephson junctions\cite{Hilgenkamp2003} and superconducting rings \cite{Davidovic1996}. The Josephson junction network is often described in terms of the $XY$ model\cite{Kolahchi1991}, with Hamiltonian $H=-J\sum_{\left\langle ij\right\rangle}cos\left(\theta_i-\theta_j\right)$ describing the coupling between phases $\theta_{ij}$. Here the summation is taken over pairs of nearest neighbours in the rings that are connected by Josephson junctions. A harmonic approximation\cite{Kusmartsev1998} so that the Josephson current is, $I=I_C sin\left(\theta_i-\theta_j\right)\approx I_C \left(\theta_i-\theta_j\right)$. It is assumed that the critical current $I_C$ of each Josephson junction is the same. When the number of Josephson junctions in a loop enables the variation in the phase to be small\cite{Kusmartsev1998} the harmonic approximation is applicable. That is to say, if the number of junctions in the plaquette is not too small then the phase dropped across each junction will be small enough that a good approximation is found. As a result, supercurrents are emphasised over phase differences in this analysis. These supercurrents circulate around the loops of the Josephson array. Using flux quantisation the current distributions for any vortex configuration is attainable. This allows the calculation of the ground state energies as well as those of the excited states in the system. The method of analysis closely mirrors Kirchoff's laws for analysing currents. Kirchoff's current conservation law states that the sum of the currents out of any vertex is zero. We now use these methods for the frustrated triangular lattice shown in Fig.\ref{Fig:triangles}. The currents are taken to flow in the clockwise sense. The self-inductance, $L$, is calculated for each side of a plaquettes and the total self-energy of the stack of triangular loops is a summation of these terms multiplied by the currents in each side. The derivation of the self-inductance is given in the supporting information. The self-inductance is written as,
\be
L_{polygon}=\frac{m~\mu_0}{4\pi}\left\{\left(D-a\right)sinh^{-1}\left(\frac{D-a}{a}\right)-a~sinh^{-1}\left(1\right)+\sqrt{2a}-\sqrt{a^2+\left(D-a\right)^2}\right\}
\label{eqn:inductance}
\ee
where $D$ is the length of one of the sides of a plaquette, $a$ is the leg-width, and $m$ is the number of legs in the polygon geometry of the $\pi$-ring (e.g. for a triangular plaquette with $L_\Delta$, m=3). In Fig.\ref{Fig:triangles} the array is composed of four triangular Josephson plaquettes. Each of these has a Josephson junction in each of its three sides. The plaquettes in this system contain the minimum number of Josephson junctions that are allowed by the method. First of all we show the results for a system of ordinary plaquettes in an applied magnetic field. The supercurrents are found from the following relationships,
\be
\left(\begin{array}{cccc} 3&0&0&-1\\ 0&3&0&-1 \\0&0&3&-1\\-1&-1&-1&3\end{array} \right) \left(\begin{array}{cccc} I_1\\I_2 \\I_3\\I_4\end{array} \right)=2\pi\left(\begin{array}{cccc} n_1-f\\n_2-f \\n_3-f\\n_4-f\end{array} \right)
\label{eqn:currents}
\ee
or in general $\mathbf{M}$ an $N_p \times N_p$ 
\be
\left(\begin{array}{ccccc} N_p&0&0&\dots&-1\\ 0&N_p&0&\dots&-1 \\0&0&N_p&\dots&-1\\ \vdots&\vdots&\vdots&\ddots&\vdots \\ -1&-1&-1&\dots&N_p  \end{array} \right) \left(\begin{array}{cccc} I_1\\I_2 \\I_3\\I_4\end{array} \right)=2\pi\left(\begin{array}{cccc} n_1-f\\n_2-f \\n_3-f\\n_4-f\end{array} \right)
\label{eqn:currents}
\ee
Note that for plaquettes that contain  $\pi$-junctions a $-1/2$ term is included in each of the $n_i-f$  terms on the right hand side of Eq.(\ref{eqn:currents}).  Here $n_i$ is the number of vortices in a plaquette with $i=1, ...,N_p$ and $N_p$ is the number of plaquettes. The parameter $f$ is an applied flux through the $i-th$ plaquette or a spontaneous flux when there exists a  $\pi$-junction. The solution of Eq.(\ref{eqn:currents}) for a triangular array gives,
\be
\begin{array}{cccc}
I_1=\frac{\pi}{27}\left(21\left(n_1-f\right)+3\left(n_2-f\right)+3\left(n_3-f\right)+9\left(n_4-f\right)\right),\\
I_2=\frac{\pi}{27}\left(3\left(n_1-f\right)+21\left(n_2-f\right)+3\left(n_3-f\right)+9\left(n_4-f\right)\right),\\
I_3=\frac{\pi}{27}\left(9\left(n_1-f\right)+9\left(n_2-f\right)+9\left(n_3-f\right)+27\left(n_4-f\right)\right),\\
I_4=\frac{\pi}{27}\left(3\left(n_1-f\right)+3\left(n_2-f\right)+21\left(n_3-f\right)+9\left(n_4-f\right)\right).
\label{eqn:In}
\end{array}
\ee
To calculate the energy of the system we write, 
\be
Energy=\frac{1}{2}\left(1-\frac{2LI_c^2}{E_J}\right)\sum_{i=1}I_i^2
\label{eqn:energy}
\ee
For the stack of triangular plaquettes of Fig.\ref{Fig:triangles} this becomes,
\be
Energy=\frac{1}{2}\left(1-\frac{2LI_c^2}{E_J}\right)\left[4\left(I_1^2+I_2^2+I_3^2\right)+6I_4^2-4\left(I_1I_4+I_2I_4+I_3I_4\right)\right]
\label{eqn:energytriangle}
\ee
In the following we write the Josephson energy equal to $\Phi_0 I_c/2 \pi$, the length of a plaquette leg as $D=45\mu m$, the leg width as $2a=15\mu m$, and the critical current as $I_c=1500aD[A]$. With these parameters and using Eq.(\ref{eqn:inductance}), the self-inductance is $L_\Delta=16pH$ and the critical current is $I_c=0.5\mu A$. Mutual inductance within the triangular array structure is expected to be insignificantly small. This conclusion is brought about by the fact that the magnetic self-energy seems to be proportionately very small in comparison to the Josephson energy (approximately $5\%$ of the size in the cases considered here). This will remain the case unless the loops become much greater in area. The additional energies will change the shape of the energy curves that are plotted as a function of the applied flux (in the case of ordinary Josephson junction plaquettes) or the spontaneous flux generated in a $\pi$-ring. However, the vortex distribution will probably be unaffected.  
\begin{figure}[!htp]
\begin{center}
\includegraphics[width=8.5cm,keepaspectratio]{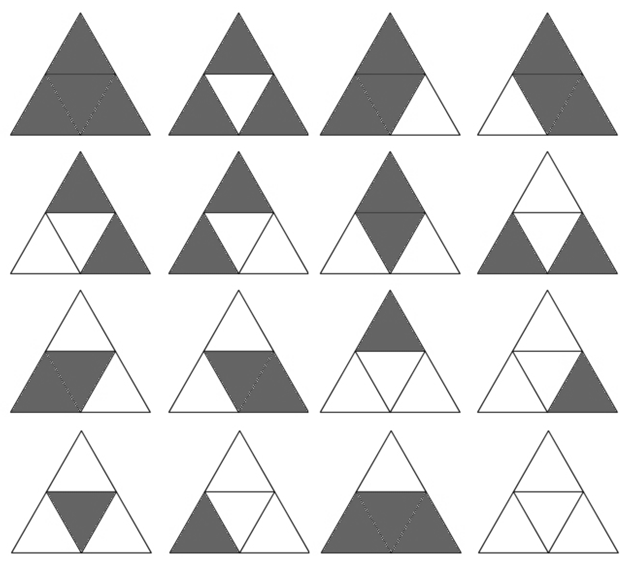}
\end{center}
\caption{The triangular networks exhibiting all configurations that represent the presence or lack of a vortex in each plaquette. Dark colouring indicates a vortex.}
\label{Fig:distribution}   
\end{figure}
All the different vortex configurations for the triangular array are shown in Fig.\ref{Fig:distribution} for the case of positive applied flux and $n=0$ or $1$. The dark coloured plaquettes depict the situation in which there is a vortex present and the unfilled ones are the empty loops. This shows the potential for logic or memory devices in such systems. In this case there are $2^4$ possible outcomes. Using this method we find the optimal configuration for the vortices within the system and hence a design ethic for quantum bits. The energy as a function of the flux (in units of the elementary flux quantum) for the various vortex patterns in the pyramid is shown in Fig.\ref{Fig:evsftriangular}. 
This plot is for the scenario whereby there are only ordinary junctions in the system and there is an applied magnetic field. In this case the energies are defined as a function of the orientation of the trapped flux: $n=0, \pm 1$. In Fig.\ref{Fig:evsftriangular} the energies and the corresponding absence or existence of a vortex gives a series of degeneracies. Each parabolic curve is related to $n_1$, $n_2$, $n_3$, $n_4$ vortices trapped by plaquettes $1-4$, respectively. Curves marked by an $A$ represent the ground state energies, with minima at $\Phi=0, \pm \Phi_0$ and $n_1=n_2=n_3=n_4=0$; $n_1=n_2=n_3=n_4=\pm 1$. Taking the region concerning positive values of quantum numbers, i.e. $0\leq f \leq 1$, which is exactly mirrored when $f\leq 0$, and increasing the flux leads to the appearance of single vortex states: i.e. curve $B$ that is related to $\left\{n_1,n_2,n_3,n_4\right\}=\left\{0,1,0,0\right\}$ or $\left\{1,0,0,0\right\}$ or $\left\{0,0,1,0\right\}$. The minimum energy of the parabola associated with these vortex configurations occurs at $f\approx 0.2$. The parabolic curve for a single vortex existing in plaquette $3$ (the central one) with no trapped fluxes in the outer three plaquettes, i.e. $\left\{0,1,0,0\right\}$, is denoted in Fig.\ref{Fig:evsftriangular} as $C$. At $f\approx 0.4$ the minimum of this condition exists. At $f\approx 0.6$ the ground state for the three outer plaquettes having vortex states is seen, i.e. curve $D$ with $\left\{1,1,1,0\right\}$. When the magnetic flux reaches $f\approx 0.8$ and when the trapping of the flux results in the scenario of $\left\{n_1,n_2,n_3,n_4\right\}=\left\{1,0,1,1\right\}$ or $\left\{0,1,1,1\right\}$ or $\left\{1,1,0,1\right\}$, the minimum of curve $E$ emerges. Again, when $f\approx 0.8$ an excited state occurs for the arrangement of vortices $\left\{1,0,0,1\right\}$, $\left\{0,1,0,1\right\}$ or $\left\{0,0,1,1\right\}$. Whenever two vortices exist in this manner in the pyramid system a higher energy state comes about through the formation of an energy repulsion between the vortices. A dimer is formed. Another excited two vortex state exists at curve $G$, $\left\{1,0,1,0\right\}$, $\left\{1,1,0,0\right\}$, or $\left\{0,1,1,0\right\}$ with a minimum energy at $f\approx 0.6$. Introducing a  $\pi$-junction to each plaquette has the effect of moving the parabolic curves of Fig.\ref{Fig:evsftriangular} by half a flux quantum. This is shown in Fig.\ref{Fig:evsftriangularpi}. 
This case, with  $\pi$-junctions included, does not need to have an external flux introduced to the system as it can naturally generate spontaneous flux - this is very important because one can then eradicate a lot of extraneous control circuitry \cite{ForresterSciRep2015}. The $\pi$-rings are therefore ideal candidates for making superconducting metamaterials \cite{fractals} as well as quantum information devices. In this discussion we will continue to develop the idea of having a central plaquette that is surrounded by exact replicas as nearest neighbours that offer topological protection. Figure (\ref{Fig:spinstar}) demonstrates a central square plaquette adjoined with four identical plaquettes on each of its vertices.  The energies are again calculated using the quasi-classical method.
These arrangements of Josephson junctions form the foundations for creating large-scale arrays with a basis for creating a layout for quantum computation. One popular design, first proposed for a two junction ring\cite{Yamashita2005}, will incorporate a ferromagnetic material as the interlayer between the superconducting islands for the $\pi$-junctions along with normal Josephson junctions with insulating barriers. The ferromagnetic layers can be metallic or, novelly, an insulator\cite{Kawabata2010}. The ferromagnetic insulators are proposed for creating  $\pi$-junction devices because they avoid the problem of strong dissipation, that is present for ferromagnetic metal systems, due to low energy quaiparticle excitations\cite{Kawabata2010}. Remarkably, in superconductor/ferromagnetic insulator/superconductor structures the ground state changes between the $0$ and $\pi$ states as the ferro-insulator increases by a single atomic layer. The existence of a $\pi$-shift is a function of the number of layers: for an even number there are positive critical currents, whereas for an odd number they are negative. Thus, there are a range of possibilities for creating $S/F/S$  structures, from high $T_c$ superconductors separated by atomic layers of, for example, $La_2BaCuO_5$, to more conventional interlayers of ferromagnetic metals that give a proximity induced $\pi$-transition. The added benefit of using the  $\pi$-states is that the multilevel energy system is spontaneously generated and consequently protected from the decoherence induced by the fluctuation of an applied magnetic field. However, one may require to use external electromagnetic fields to manipulate the system, so again the issue of protecting the qubit states arises. Thus, building the spin star kind of array can result in systems that function to do so.  

At this point it is time to also point out that schemes do exist with completely different architectures to those proposed for standard quantum computation. Quantum logic gates form the foundation for one methodology. This approach, whilst enormously promising, would seem to be many years from creating large scale quantum computing. Currently, a Canadian company called D-wave are marketing a quantum computer designed around a different concept using quantum annealing. However, there is a debate going on as to whether quantum annealing offers any speed up on classical computers. Quantum annealing searches for the lowest energy configuration of the system [Boixo 2014] and so the techniques we have described here are equally relevant for qubit architectures based on quantum gating, annealing [Boixo 2014] or adiabatic quantum computing [Deng 2013]. The D-wave machines are very much prototypes for the concept and it is therefore prudent to keep investigating innovative new techniques and methods of analysis to  avoid the problem of noise in the system. 

Another very promising form of quantum computation using superconducting systems is that of the topologically protected system\cite{IoffeFeigelman2002,IoffeFeigelman2003}, of which the spin-star is a subset. Josephson junction arrays imitate dimers, as we saw for the pyramid structure, when composed of triangular or kagome networks.A Josephson junction network with protected ground state degeneracies is one of the most promising long term solutions to the issues surrounding decoherence. Also, the vortices in the system may be propagated through the topological sectors by a slow, adiabatic, change of the flux towards a half integer value. In this way many of the principles guiding the design of functional Josephson arrays for quantum computing can be amalgamated to produce an optimally effective solution - topological degeneracy of ground states in a protected subspace, like that proposed by [Kitaev 2003 ]  with adiabatic evolution or quantum gates. Bose and Hutton investigated a central spin that interacted with a set of peripheral spins at low temperature [Bose/Hutton 2004]. That system is directly analogous to the one we are designing here, except instead of spins we take the quantum numbers associated with the trapped fluxes as the pertinent states. The central vortex will entangle with the other vortices at low temperature with a dependence on the total number of vortices in the system.  

%\bibliography{DMFtopologicalqubitsBIB}

\begin{thebibliography}{10}

\bibitem{Davidovic1996}
D.~Davidovi\`c, S.~Kumar, D.~H. Reich, J.~Siegel, S.~B. Field, R.~C. Tiberio,
  R.~Hey, and K.~Ploog.
\newblock {C}orrelations and disorder in arrays of magnetically coupled
  superconducting rings.
\newblock {\em { P}hys. {R}ev. {L}ett.}, 76(5):815--818, 1996.

\bibitem{fractals}
M.~Forrester, K.~Kurten, and F.~Kusmartsev.
\newblock Fractal metamaterials composed of electrically isolated $\pi$-rings.
\newblock {\em Sci. Lett. J.}, 4:133, 2015.

\bibitem{ForresterSciRep2015}
M.~Forrester and F.~Kusmartsev.
\newblock Whispering galleries and the control of artificial atoms.
\newblock {\em Scientific reports}, 6:25084, 2016.

\bibitem{Frolov2008}
S.~M. Frolov, M.~J.~A. Stoutimore, T.~A. Crane, D.~J.~Van Harlingen, V.~A.
  Oboznov, V.~V. Ryazanov, A.~Ruosi, C.~Granata, and M.~Russo.
\newblock {I}maging spontaneous currents in superconducting arrays of $\pi$
  -junctions.
\newblock {\em {N}ature {P}hysics}, 4:32--36, 2008.

\bibitem{Hilgenkamp2003}
H.~Hilgenkamp, Ariando, H.~J.~H. Smilde, D.~H.~A. Blank, G.~Rijnders,
  H.~Rogalla, J.~R. Kirtley, and C.~C. Tsuei.
\newblock {O}rdering and manipulation of the magnetic moments in large-scale
  superconducting $\pi$-loop arrays.
\newblock {\em {N}ature}, 422:50--53, 2003.

\bibitem{IoffeFeigelman2003}
L.~B. Ioffe and M.~V. Feigel'man.
\newblock {R}ealization of topologically protected quantum bits in a josephson
  junction array.
\newblock {\em {P}hys. {U}sp.}, 46:759--764, 2003.

\bibitem{IoffeFeigelman2002}
L.~B. Ioffe, M.~V. Feigel'man, A.~Ioselevich, D.~Ivanov, M.~Troyer, and
  G.~Blatter.
\newblock {T}opologically protected quantum bits using josephson junction
  arrays.
\newblock {\em {N}ature}, 415:503--506, 2002.

\bibitem{Kolahchi1991}
M.~R. Kolahchi and J.~P. Straley.
\newblock {G}round state of the uniformly frustrated two-dimensional $xy$ model
  near $f=1/2$.
\newblock {\em { P}hys. {R}ev. {B}.}, 43:7651--7654, 1991.

\bibitem{Kusmartsev1998}
F.~V. Kusmartsev, D.M. Faruque, and D.~I. Khomskii.
\newblock {Q}uasiclassical method for a josephson network in a magnetic field.
\newblock {\em {P}hysics Letters A}, 249:541--554, 1998.

\bibitem{Yamashita2005}
T.~Yamashita, K.~Tanikawa, S.~Takahashi, and S.~Maekawa.
\newblock {S}uperconducting $\pi$ qubit with a ferromagnetic josephson
  junction.
\newblock {\em { P}hys. {R}ev. {L}ett.}, 95:097001, 2005.

\end{thebibliography}

\end{document}